\documentclass[12pt,a4paper]{article}

\usepackage{graphicx}  
\usepackage{xspace}
\usepackage{color}
\usepackage{colortbl}
\usepackage{multirow}
\usepackage{amsmath}
\usepackage{ifthen} 

\usepackage{graphpap}
\usepackage{rotating}
\usepackage{cancel}
\usepackage{hyperref}
\usepackage[all]{hypcap}

\newboolean{pdflatex}
\setboolean{pdflatex}{true} 
\setboolean{pdflatex}{true} 

\newboolean{uprightparticles}
\setboolean{uprightparticles}{true} 
\newboolean{articletitles}
\setboolean{articletitles}{true} 
\usepackage{amssymb}
\usepackage{amsfonts}
\usepackage{upgreek}

\ifthenelse{\boolean{uprightparticles}}%
{

 \def\PDelta      {\ensuremath{\Delta}\xspace}                 
 \def\PXi      {\ensuremath{\Xi}\xspace}                 
 \def\PLambda      {\ensuremath{\Lambda}\xspace}                 
 \def\PSigma      {\ensuremath{\Sigma}\xspace}                 
 \def\POmega      {\ensuremath{\Omega}\xspace}                 
 \def\PUpsilon      {\ensuremath{\Upsilon}\xspace}                 
 
 \def\PB      {\ensuremath{\mathrm{B}}\xspace}                 
                  
 \def\PD      {\ensuremath{\mathrm{D}}\xspace}

 \def\PK      {\ensuremath{\mathrm{K}}\xspace}

 \def\Pi      {\ensuremath{\mathrm{i}}\xspace}

}
{

 \mathchardef\PDelta="7101
 \mathchardef\PXi="7104
 \mathchardef\PLambda="7103
 \mathchardef\PSigma="7106
 \mathchardef\POmega="710A
 \mathchardef\PUpsilon="7107
                  
 \def\PB      {\ensuremath{B}\xspace}                 
                  
 \def\PD      {\ensuremath{D}\xspace}

 \def\PK      {\ensuremath{K}\xspace}

 \def\Pi      {\ensuremath{i}\xspace}

}

\def\kaon  {\ensuremath{\PK}\xspace}
  \def\Kbar  {\kern 0.2em\overline{\kern -0.2em \PK}{}\xspace}

\def\Kz    {\ensuremath{\kaon^0}\xspace}
\def\Kzb   {\ensuremath{\Kbar^0}\xspace}
\def\KzKzb {\ensuremath{\Kz \kern -0.16em \Kzb}\xspace}
\def\Kp    {\ensuremath{\kaon^+}\xspace}
\def\Km    {\ensuremath{\kaon^-}\xspace}

\def\KpKm  {\ensuremath{\Kp \kern -0.16em \Km}\xspace}

  \def\Dbar    {\kern 0.2em\overline{\kern -0.2em \PD}{}\xspace}
\def\D       {\ensuremath{\PD}\xspace}

\def\Dz      {\ensuremath{\D^0}\xspace}
\def\Dzb     {\ensuremath{\Dbar^0}\xspace}
\def\DzDzb   {\ensuremath{\Dz {\kern -0.16em \Dzb}}\xspace}
\def\Dp      {\ensuremath{\D^+}\xspace}
\def\Dm      {\ensuremath{\D^-}\xspace}

\def\DpDm    {\ensuremath{\Dp {\kern -0.16em \Dm}}\xspace}

  \def\Bbar    {\kern 0.18em\overline{\kern -0.18em \PB}{}\xspace}

  \def\Y#1S{\ensuremath{\PUpsilon{(#1S)}}\xspace}

\newcommand{\tev}{\ensuremath{\mathrm{\,Te\kern -0.1em V}}\xspace}
\newcommand{\gev}{\ensuremath{\mathrm{\,Ge\kern -0.1em V}}\xspace}
\newcommand{\mev}{\ensuremath{\mathrm{\,Me\kern -0.1em V}}\xspace}
\newcommand{\kev}{\ensuremath{\mathrm{\,ke\kern -0.1em V}}\xspace}
\newcommand{\ev}{\ensuremath{\mathrm{\,e\kern -0.1em V}}\xspace}
\newcommand{\gevc}{\ensuremath{{\mathrm{\,Ge\kern -0.1em V\!/}c}}\xspace}
\newcommand{\mevc}{\ensuremath{{\mathrm{\,Me\kern -0.1em V\!/}c}}\xspace}
\newcommand{\gevcc}{\ensuremath{{\mathrm{\,Ge\kern -0.1em V\!/}c^2}}\xspace}
\newcommand{\gevgevcccc}{\ensuremath{{\mathrm{\,Ge\kern -0.1em V^2\!/}c^4}}\xspace}
\newcommand{\mevcc}{\ensuremath{{\mathrm{\,Me\kern -0.1em V\!/}c^2}}\xspace}

\def\nb {\ensuremath{\rm \,nb}\xspace}

\def\invpb {\ensuremath{\mbox{\,pb}^{-1}}\xspace}

\def\to                 {\ensuremath{\rightarrow}\xspace}

\def\gsim{{~\raise.15em\hbox{$>$}\kern-.85em
          \lower.35em\hbox{$\sim$}~}\xspace}
\def\lsim{{~\raise.15em\hbox{$<$}\kern-.85em
          \lower.35em\hbox{$\sim$}~}\xspace}

\def\AT#1     {\ensuremath{A_T^{#1}}\xspace}           

\def\C#1      {\ensuremath{\mathcal{C}_{#1}}}                       
\def\Cp#1     {\ensuremath{\mathcal{C}_{#1}^{'}}}                    
\def\Ceff#1   {\ensuremath{\mathcal{C}_{#1}^{\mathrm{(eff)}}}}        
\def\Cpeff#1  {\ensuremath{\mathcal{C}_{#1}^{'\mathrm{(eff)}}}}       
\def\Ope#1    {\ensuremath{\mathcal{O}_{#1}}}                       
\def\Opep#1   {\ensuremath{\mathcal{O}_{#1}^{'}}}                    

\newcommand{\theresult}{5.4} 
\newcommand{\thestat}{1.3}   
\newcommand{\thesyst}{0.8}   

\newcommand{\X}{\ensuremath{X(3872)}\xspace}

\usepackage{mciteplus}

\begin{document}

\renewcommand{\thefootnote}{\fnsymbol{footnote}}
\setcounter{footnote}{1}

\begin{titlepage}
\pagenumbering{roman}

\vspace*{-1.5cm}
\centerline{\large EUROPEAN ORGANIZATION FOR NUCLEAR RESEARCH (CERN)}
\vspace*{1.5cm}
\hspace*{-0.5cm}
\begin{tabular*}{\linewidth}{lc@{\extracolsep{\fill}}r}
\ifthenelse{\boolean{pdflatex}}
{\vspace*{-2.7cm}\mbox{\!\!\!\includegraphics[width=.14\textwidth]{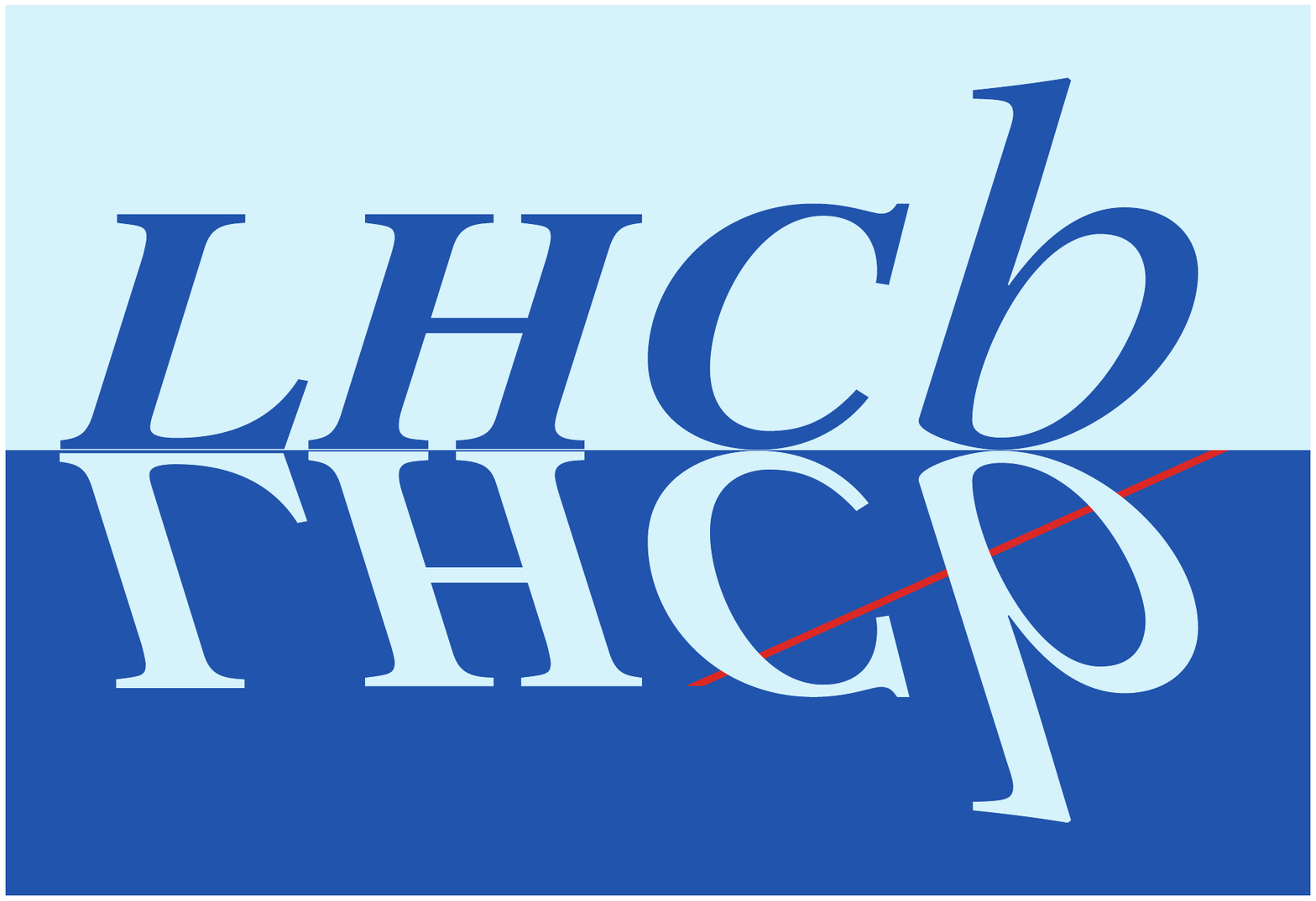}} & &}%
{\vspace*{-1.2cm}\mbox{\!\!\!\includegraphics[width=.12\textwidth]{lhcb-logo.eps}} & &}%
\\
 & & LHCb-PAPER-2011-034 \\  
 & & CERN-PH-EP-2011-216 \\  
 & & December 20, 2011; rev.\ March 22, 2012 \\
 & &        \\ 

\end{tabular*}
\vspace*{4.0cm}

{\bf\boldmath\huge
\begin{center}
  Observation of \X production in $pp$ collisions at $\sqrt{s}=7\tev$
\end{center}
}
\vspace*{1.0cm}

\begin{center}
The LHCb collaboration
\footnote{Authors are listed on the following pages.}
\end{center}

\begin{abstract}
  \noindent
Using 34.7\invpb of data collected with the LHCb detector, the
inclusive production of the \X meson in $pp$ collisions
at $\sqrt{s}=7\tev$ is observed for the first time. 
Candidates are selected in the $\X \to J/\psi \pi^+ \pi^-$ decay mode, and used to measure 
\begin{align*}
\sigma(pp\to \X+{\rm anything}) \, \mathcal{B}(\X \to J/\psi&\pi^+\pi^-) = \\ 
& \theresult \pm \thestat \,{\rm (stat)} \pm \thesyst \,{\rm (syst)}\nb \,,
\end{align*}
where $\sigma(pp\to \X+{\rm anything})$ is the inclusive production cross-section of \X mesons 
with rapidity in the range $2.5-4.5$ and transverse momentum 
in the range $5-20\gevc$. In addition the masses of both the \X and $\psi(2S)$ mesons, 
reconstructed in the $ J/\psi \pi^+ \pi^-$ final state, are measured to be
\begin{eqnarray*}
m_{\X} &=& 3871.95 \pm 0.48 \,({\rm stat}) \pm 0.12 \,({\rm syst})\mevcc ~\mbox{and} \\
m_{\psi(2S)} &=& 3686.12\pm 0.06 \,({\rm stat}) \pm 0.10 \,({\rm syst})\mevcc \,.
\end{eqnarray*} 
\end{abstract}
\vspace*{0.50cm}
\vspace{\fill}
\begin{center}
{\it (Published in Eur.\ Phys.\ J.\ C 72 (2012) 1972)}
\end{center}

\end{titlepage}


\renewcommand{\thefootnote}{\arabic{footnote}}
\setcounter{footnote}{0}
\newpage
\setcounter{page}{2}
\mbox{~}
\newpage

\begin{flushleft}
\centerline{\large\bf The LHCb Collaboration} 
\vspace{2ex}
\small 
R.~Aaij$^{23}$, 
C.~Abellan~Beteta$^{35,n}$, 
B.~Adeva$^{36}$, 
M.~Adinolfi$^{42}$, 
C.~Adrover$^{6}$, 
A.~Affolder$^{48}$, 
Z.~Ajaltouni$^{5}$, 
J.~Albrecht$^{37}$, 
F.~Alessio$^{37}$, 
M.~Alexander$^{47}$
% ...

{\it
\footnotesize 
% ...
$ ^{5}$Clermont Universit\'{e}, Universit\'{e} Blaise Pascal, CNRS/IN2P3, LPC, Clermont-Ferrand, France\\
$ ^{6}$CPPM, Aix-Marseille Universit\'{e}, CNRS/IN2P3, Marseille, France\\
% ...
$ ^{23}$Nikhef National Institute for Subatomic Physics, Amsterdam, The Netherlands\\
% ...
$ ^{35}$Universitat de Barcelona, Barcelona, Spain\\
$ ^{36}$Universidad de Santiago de Compostela, Santiago de Compostela, Spain\\
$ ^{37}$European Organization for Nuclear Research (CERN), Geneva, Switzerland\\
% ...
$ ^{42}$H.H. Wills Physics Laboratory, University of Bristol, Bristol, United Kingdom\\
% ...
$ ^{47}$School of Physics and Astronomy, University of Glasgow, Glasgow, United Kingdom\\
$ ^{48}$Oliver Lodge Laboratory, University of Liverpool, Liverpool, United Kingdom\\
% ...
$ ^{n}$LIFAELS, La Salle, Universitat Ramon Llull, Barcelona, Spain\\
% ...
}
\bigskip
\end{flushleft}

\cleardoublepage


\pagestyle{plain} 
\setcounter{page}{1}
\pagenumbering{arabic}

\section{Introduction}
\label{sec:Introduction}

The \X particle was discovered in 2003 by the Belle
collaboration in the $B^{\pm} \to \X K^{\pm}$, $\X \to  J/\psi \pi^{+} \pi^{-}$
decay chain~\cite{Choi:2003ue}. Its existence was confirmed by the
CDF~\cite{CDFPhysRevLett.93.072001}, 
D\O~\cite{D0Abazov:2004kp} and BaBar~\cite{BaBarPhysRevD.71.071103} collaborations. 
The discovery of the \X particle and the subsequent observation of several other new states 
in the mass range $3.9-4.7~\gevcc$ have led 
to a resurgence of interest in exotic meson spectroscopy~\cite{Swanson2006243}.

Several properties of the \X have been determined, in particular its mass
~\cite{Aaltonen:2009vj, Choi:2011fc, Aubert:2008gu} 
and the dipion mass spectrum in the decay $\X\to J/\psi\pi^+\pi^-$ 
\cite{Abulencia:2005zc,Choi:2011fc}, but its quantum numbers, 
which have been constrained to be either 
$J^{PC} = 2^{-+}$ or $1^{++}$~\cite{Abulencia:2006ma}, are still 
not established. Despite a large experimental effort,
the nature of this new state is still uncertain and
several models have been proposed to describe it.
The \X could be a conventional charmonium state,
with one candidate being the  $\eta_{c2}(1D)$
meson~\cite{Swanson2006243}. However, the mass of this state is predicted to be
far below the observed \X mass. Given the proximity of the \X
mass to the $D^{*0}\bar{D}^0$ threshold, another possibility
is that the \X is a loosely bound 
$D^{*0}\bar{D}^0$ `molecule',
i.e.\ a $((u\overline{c})(c\overline{u}))$
system~\cite{Swanson2006243}. For this interpretation to be valid the
mass of the \X should be less than the sum of $D^{*0}$ and ${D}^0$
masses. A further, more exotic, possibility is that the \X is a tetraquark state~\cite{Maiani:2004vq}. 

Measurements of $X(3872)$ production at hadron colliders, where most of the production
is prompt rather than from $b$-hadron decays, may shed light
on the nature of this particle. In particular, it has been discussed whether 
or not the possible molecular nature of the \X is compatible with the production
rate observed at the Tevatron~\cite{Bignamini:2009sk,Artoisenet:2009wk}.
Predictions for \X production at the LHC have also been published~\cite{Artoisenet:2009wk}. 

This paper reports an observation of \X production in $pp$ 
collisions at $\sqrt{s}~=7~\tev$ using 
an integrated luminosity of 34.7~\invpb 
collected by the LHCb experiment.
The $\X\to J/\psi\pi^+\pi^-$ selection is optimized on the similar
but more abundant $\psi(2S) \to J/\psi\pi^+\pi^-$ decay.
The observed \X signal is used to measure both the \X mass and the production rate
from all sources including $b$-hadron decays, 
 i.e.\ the absolute inclusive \X 
production cross-section in the detector acceptance
multiplied by the $\X\to J/\psi\pi^+\pi^-$ branching fraction.

\section{The LHCb spectrometer and data sample}
\label{sec:detector}

The LHCb detector is a forward spectrometer~\cite{LHCbDetector} 
at the Large Hadron Collider (LHC).
It provides reconstruction of charged particles in the
pseudorapidity range $2<\eta<5$. 
The detector elements are 
placed along the LHC beam line starting with the vertex detector (VELO), a
silicon strip device that surrounds the proton-proton interaction
region. It is used to reconstruct both the interaction vertices
and the decay vertices of long-lived hadrons.
It also contributes to the measurement of track
momenta, along with a large area silicon strip detector located 
upstream of a dipole magnet 
and a combination of silicon strip detectors and straw 
drift-tubes placed downstream. 
The magnet has a bending power of about 4~Tm. 
The combined tracking system has 
a momentum resolution $\delta p/p$ that varies from 0.4\% 
at 5~\gevc to 0.6\% at 100~\gevc.
Two ring imaging Cherenkov (RICH) detectors are used to identify charged hadrons. 
The detector is completed by electromagnetic calorimeters for photon and
electron identification, a hadron calorimeter, and a muon 
system consisting of alternating layers of iron and multi-wire proportional chambers. 
The trigger consists of a hardware stage, based on information from the calorimeter and muon systems,
followed by a software stage which applies a full event reconstruction. 

The cross-section analysis described in this paper is based on a data sample collected in 2010, 
exclusively using events that passed dedicated $J/\psi$ trigger algorithms. 
These algorithms selected 
a pair of oppositely-charged muon candidates, where either one of the muons had
a transverse momentum  $p_{\rm T}$ larger than 1.8~\gevc or one of the two muons
had $p_{\rm T}> 0.56$~\gevc and the other $p_{\rm T} > 0.48$~\gevc. 
The pair of muons was required to originate from a common vertex and have an invariant mass in a wide window around the $J/\psi$ mass. 
The \X mass measurement also uses events triggered with other algorithms, such as single-muon triggers. 
To avoid domination of the trigger CPU time by a few events with high occupancy, 
a set of cuts 
was applied on the hit multiplicity of each sub-detector used 
by the pattern recognition algorithms. 
These cuts reject high-multiplicity events with a large number 
of $pp$~interactions. 

The accuracy of the \X mass measurement relies on the calibration
of the tracking system~\cite{bmass}.
The spatial alignment of
the tracking detectors, as well as the  calibration of the momentum scale, are
based on the $J/\psi \to \mu^+\mu^-$ mass peak.
This was carried out in seven time periods corresponding to known changes in
the detector running conditions.
The procedure takes into account the effects of QED radiative 
corrections which are important in this decay.

The analysis uses fully simulated samples based on 
the {\sc{Pythia}}~6.4 generator~\cite{pythia} configured 
with the parameters detailed in Ref.~\cite{belyaev}. 
The {\sc{EvtGen}}~\cite{evtgen}, {\sc{Photos}}~\cite{photos}
and {\sc{Geant4}}~\cite{geant} packages are used 
to describe the decays of unstable particles, model QED radiative corrections
and simulate interactions in the detector, respectively. 
The $\X\to J/\psi\pi^+\pi^-$ Monte Carlo events are generated assuming
that the $\rho$ resonance dominates the dipion mass spectrum, as
established by the CDF~\cite{Abulencia:2005zc} and Belle~\cite{Choi:2011fc} data. 

\section{Event selection}
\label{sec::selection}

To isolate the \X signal, tight cuts are needed to reduce combinatorial background
where a correctly reconstructed $J/\psi$ meson is combined with a random 
$\pi^+\pi^-$ pair from the primary
$pp$ interaction. The cuts are defined using reconstructed $\psi(2S) \to J/\psi \pi^+\pi^-$ 
decays, as well as `same-sign pion' candidates satisfying the same criteria as used for the
\X and $\psi(2S)$ selection but where the two pions have the same electric charge.  The Kullback-Leibler (KL) distance~\cite{kl,*kl2,*kl3}
is used to suppress duplicated particles created by the reconstruction:
if two particles have a symmetrized KL divergence less than 5000,
only that with the higher track fit quality is considered.

$J/\psi\to\mu^+\mu^-$ candidates are formed from pairs of oppositely-charged
particles identified as muons, originating from a common vertex with a $\chi^2$ per degree of freedom ($\chi^2/{\rm ndf}$) smaller than $ 20$,
and with an invariant mass in the range $3.04-3.14~\gevcc$.
The two muons are each required to have a momentum 
above $10~\gevc$ and a transverse momentum 
above $1~\gevc$. To reduce background from the decay in flight of pions
and kaons, each muon candidate is required to have a 
track fit $\chi^2/{\rm ndf}$ less than 4.
Finally $J/\psi$ candidates are required
to have a transverse momentum larger than $3.5~\gevc$.

Pairs of oppositely-charged pions are combined with $J/\psi$ candidates to build $\psi(2S)$ and \X candidates.
To reduce the 
combinatorial background, each pion candidate is required to have a transverse momentum above $0.5~\gevc$ 
and a track fit $\chi^2/{\rm ndf}$ less than 4. 
In addition, kaons are removed using the RICH information by requiring the likelihood for the kaon hypothesis to be smaller than that for the pion hypothesis.
A vertex fit is performed~\cite{Hulsbergen} that constrains the four daughter particles to originate from a common point 
and the mass of the muon pair to the nominal $J/\psi$
mass~\cite{Nakamura:2010zzi}. This fit both improves the mass
resolution and reduces the sensitivity of the result to the momentum
scale calibration. To further reduce the combinatorial background
the $\chi^2/{\rm ndf}$ of this fit is required to be less than 5.
Finally, the requirement $Q < 300~\mevcc$ 
is applied where $Q = M_{\mu\mu\pi\pi} - M_{\mu\mu} - M_{\pi\pi}$, and 
$M_{\mu\mu\pi\pi}$, $M_{\mu\mu}$ and $M_{\pi\pi}$ are 
the reconstructed masses before any mass constraint; this requirement 
removes $35\%$ of the background whilst retaining $97\%$ of the 
\X signal.

\begin{figure}[t]
\centering
\includegraphics[width=0.8\textwidth]{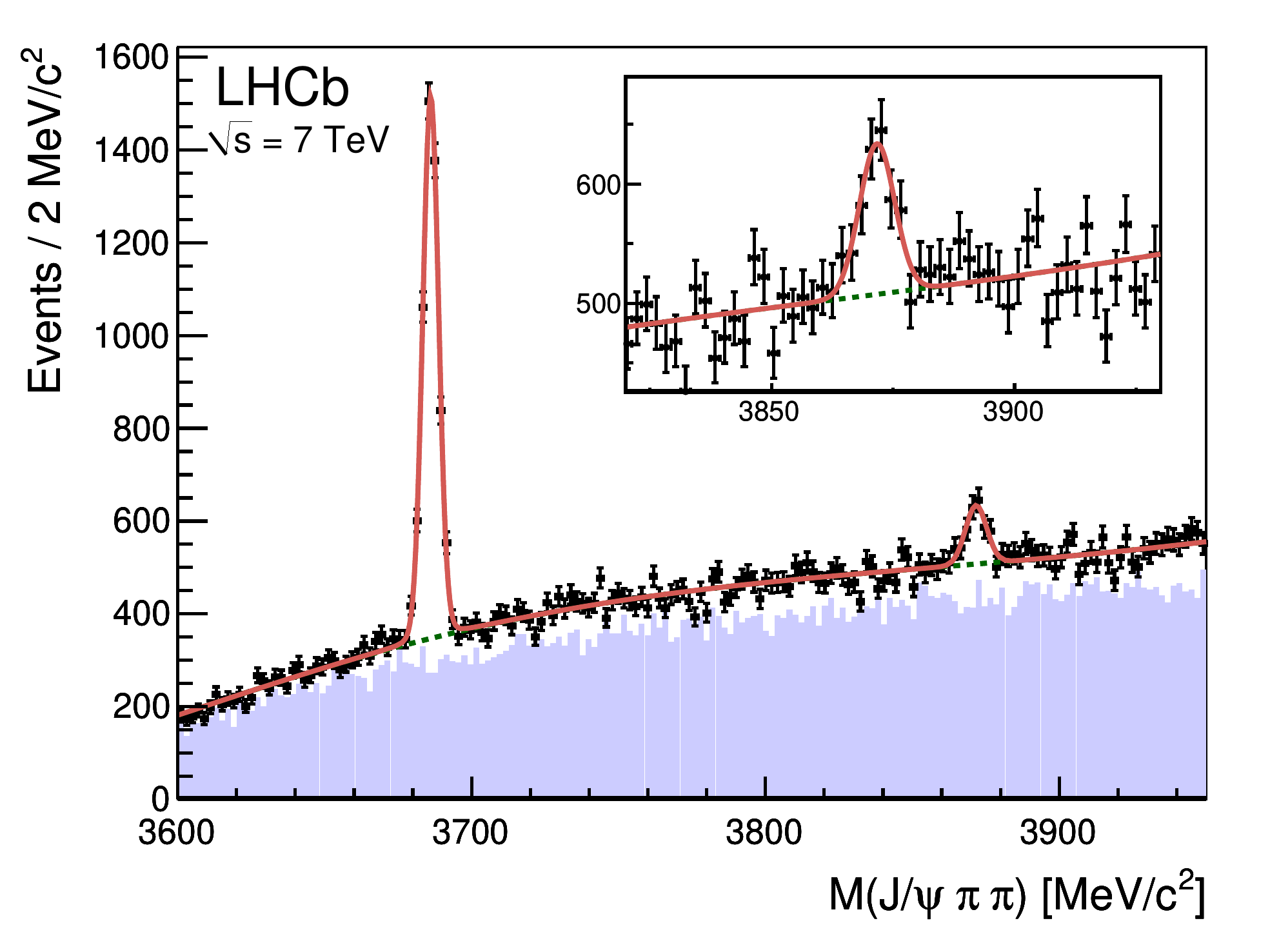} 
\caption{\small Invariant mass distribution of $J/\psi \pi^+ \pi^-$ (points with statistical error bars) and same-sign $J/\psi \pi^{\pm} \pi^{\pm}$ (filled histogram)
candidates. The curves are the result of the fit described in the
text. The inset shows a zoom of the \X region.}
\label{X3872}
\end{figure}

Figure~\ref{X3872} shows the $J/\psi \pi^+ \pi^-$ mass distribution for
the selected candidates, with clear signals for both the $\psi(2S)$ and the \X mesons,
as well as the $J/\psi \pi^{\pm} \pi^{\pm}$ mass distribution of the same-sign pion candidates.

\section{Mass measurements}
\label{sec:mass}

The masses of the $\psi(2S)$ and \X mesons are determined from an extended unbinned maximum
likelihood fit of the reconstructed $J/\psi \pi^+ \pi^-$ mass in the
interval $3.60 < M_{J/\psi \pi\pi} < 3.95~\gevcc$. The $\psi(2S)$ and \X signals are each described with a 
non-relativistic Breit-Wigner function convolved with a Gaussian resolution function. 
The intrinsic width of the $\psi(2S)$ is fixed to the PDG value,
$\Gamma_{\psi(2S)} = 0.304\mevcc$~\cite{Nakamura:2010zzi}. The Belle
collaboration recently reported~\cite{Choi:2011fc} that the \X width is
less than $1.2$~\mevcc at $90\%$ confidence level; we
fix the \X width to zero in the nominal fit. 
The ratio of the mass resolutions for the \X and the $\psi(2S)$ is
fixed to the value estimated from the simulation,
$\sigma^{\rm MC}_{\X}/\sigma^{\rm MC}_{\psi(2S)} = 1.31$.

Studies using the same-sign pion candidates show that the background shape can be 
described by the functional form $f(M) \propto (M-m_{\rm th})^{c_0} \exp(-c_1 M -c_2 M^2)$, 
where $m_{\rm th} = m_{J/\psi} + 2 m_{\pi} = 3376.05~\mevcc$~\cite{Nakamura:2010zzi} 
is the mass threshold and $c_0$, $c_1$ and $c_2$ are shape parameters. 
To improve the stability of the fit, the parameter $c_2$ is
fixed to the value obtained from the same-sign pion sample.

In total, the fit has eight free parameters: three yields ($\psi(2S)$, \X and background), 
two masses ($\psi(2S)$ and \X), one resolution parameter, and two background shape parameters. 
The correctness of the fitting procedure has been checked with simplified
Monte Carlo samples, fully simulated Monte Carlo samples, 
and samples containing a mixture of fully
simulated Monte Carlo signal events and same-sign background events taken
from the data. The fit results are shown in Fig.~\ref{X3872} and Table~\ref{tab:fitResults}.
The fit does not account for 
QED radiative corrections and hence underestimates the masses.
Using a simulation based on {\sc{Photos}}~\cite{photos} the biases
on the \X and $\psi(2S)$ masses are
found to be $-0.07 \pm 0.02 \mevcc$ and $-0.02 \pm 0.02 \mevcc$, 
respectively. The
fitted mass values are corrected for these biases and the
uncertainties propagated in the estimate of the systematic error.

\begin{table}[t]
\caption{\small Results of the fit to the $J/\psi \pi^+ \pi^-$ invariant mass distribution of Fig.~\ref{X3872}.} 
\begin{center}
\begin{tabular}{l|r@{\,}l|r@{\,}l} 
Fit parameter or derived quantity & \multicolumn{2}{c|}{$\psi(2S)$} & \multicolumn{2}{c}{\X}   \\ \hline 
Number of signal events & $3998$ & $\pm\, 83$    & $565$ & $\pm\, 62$ \\
Mass $m~[\mevcc]$       & $3686.10$ & $\pm\, 0.06$    & $3871.88$ & $\pm\, 0.48$  \\
Resolution $\sigma~[\mevcc]$   &  $2.54$ & $\pm\, 0.06$ & $3.33$ & $\pm\, 0.08$ \\  
Signal-to-noise ratio in $\pm 3\sigma$ window  &  $1.5$ & & $0.15$ & \\ 
Number of background events  &   \multicolumn{4}{c}{$73094 \pm 282$ ~~~} \\
\end{tabular}
\end{center}
\label{tab:fitResults}
\end{table}

Several other sources of systematic effects on the mass measurements are considered. For each source, 
the complete analysis is repeated (including the track fit and the momentum scale calibration when needed)
under an alternative assumption, and the observed change in the central value of the 
fitted masses relative to the nominal results assigned as a systematic
uncertainty. The dominant source of uncertainty is due to the
calibration of the momentum scale.
Based on checks performed with reconstructed signals of various mesons 
decaying into two-body final states (such as $\pi^+\pi^-$, $K^{\mp}\pi^{\pm}$ and $\mu^+\mu^-$)
a relative systematic uncertainty of 0.02\% is assigned to the momentum scale~\cite{bmass},
which translates into a 0.10 (0.08)~\mevcc uncertainty on the \X ($\psi(2S)$) mass.
After the calibration procedure with the $J/\psi \rightarrow \mu^+ \mu^-$ decay,
a $\pm 0.07\%$ variation of the momentum scale remains as a function of the particle pseudorapidity $\eta$.
To first order this effect averages out in the mass determination.
The residual impact of this variation is evaluated by parameterizing the
momentum scale as function of $\eta$ and repeating the analysis.
The systematic uncertainty associated with the momentum calibration
indirectly takes into account any effect related to the imperfect
alignment of the tracking stations. However, the alignment of the
VELO may affect the mass measurements through the
determination of the horizontal and vertical slopes of the tracks.
This is investigated by changing the track slopes
by amounts corresponding to the 0.1\% relative precision
with which the length scale along the beam axis is known~\cite{dms}.
Other small uncertainties arise
due to the limited knowledge of the \X width and the modelling of the resolution. 
The former is estimated by fixing the \X width to 0.7~\mevcc instead of zero, 
as suggested by the likelihood published by Belle~\cite{Choi:2011fc}.
The latter is estimated by fixing the ratio $\sigma_{\X}/\sigma_{\psi(2S)}$
using the covariance estimates returned by the track fit algorithm on signal
events in the data sample,
rather than using the mass resolutions from the simulation.
The effect of background modelling is estimated by performing the fit on
two large samples, one with only Monte Carlo signal events,
and one containing a mixture of Monte Carlo signal events and background
candidates obtained by combining a $J/\psi$ candidate and a same-sign
pion pair from different data events: the difference in the fitted
mass values is taken as a systematic uncertainty.
The amount of material
traversed in the tracking system by a particle is estimated to be
known to a 10\% accuracy~\cite{Aaij:2010nx}; the magnitude of the energy 
loss correction in the reconstruction is therefore varied by 10\%. 
The assigned systematic uncertainties are summarized in Table~\ref{systematictab} and combined in quadrature. 
\begin{table}[t]
\caption{\small Systematic uncertainties on the $\psi(2S)$ and \X
mass measurements.}
\begin{center}
\begin{tabular}{l|l|c|c} 
\multirow{2}{*}{Category}  & \multirow{2}{*}{Source of uncertainty}
  & \multicolumn{2}{c} {$\Delta m$ [\!$\mevcc$]}  \\ 
& & $\psi(2S)$ & \X \\ \hline 
\multirow{4}{*}{Mass fitting}
& Natural width    &  --    & 0.01 \\
& Radiative tail   &  0.02  & 0.02 \\
& Resolution & -- & 0.01 \\
& Background model & 0.02  & 0.02\\
\hline 
\multirow{2}{*}{Momentum calibration}
& Average momentum scale  &    0.08   & 0.10 \\ 
& $\eta$ dependence of momentum scale & 0.02   & 0.03  \\ 
\hline
\multirow{1}{*}{Detector description}
& Energy loss correction      & 0.05  & 0.05 \\ 
\hline 
Detector alignment
& Track slopes  & 0.01    & 0.01 \\   
\hline \hline 
Total &                       & 0.10 & 0.12 \\ 
\end{tabular}
\end{center}
\label{systematictab}
\end{table}

Systematic checks of the stability of the measured $\psi(2S)$ mass
are performed, splitting the data sample according to different
run periods or to the dipole magnet polarity, or ignoring the 
hits from the tracking station before the magnet. 
In addition, the measurement is repeated in bins of the
$p$, $p_{\rm T}$ and $Q$ values of the $\psi(2S)$ signal.
No evidence for a systematic bias is found.

\section{Determination of the production cross-section}
\label{sec:systematics}
The observed \X signal is used to measure the product of the
inclusive production cross-section $\sigma(pp\to \X+{\rm anything})$
and the branching fraction $\mathcal{B}(\X \to J/\psi\pi^+\pi^-)$, 
according to
\begin{equation}
\sigma(pp\to \X+{\rm anything}) \, \mathcal{B}(\X \to J/\psi\pi^+\pi^-) =
\frac{N^{\rm corr}_{\X}}{
\xi \, \mathcal{B}(J/\psi\to\mu^+\mu^-)
\, \mathcal{L}_{\rm int}} \,,
\label{eq:sigmaBR}
\end{equation}
where $N^{\rm corr}_{\X}$ is the efficiency-corrected signal yield,
$\xi$ is a correction factor to the simulation-derived efficiency
that accounts for known differences between data and simulation,
$\mathcal{B}(J/\psi\to\mu^+\mu^-) = (5.93 \pm 0.06)\%$~\cite{Nakamura:2010zzi} is the 
$J/\psi\to\mu^+\mu^-$ branching fraction, 
and $\mathcal{L}_{\rm int}$ is the integrated luminosity.

The absolute luminosity scale was measured at specific periods during the 2010 
data taking~\cite{luminosity-paper} using both Van der Meer scans~\cite{VanDerMeer} and a beam-gas
imaging method~\cite{BeamProfile}. The instantaneous luminosity determination 
is then based on a continuous recording of the multiplicity of tracks in the 
VELO, which has been normalized to the absolute luminosity scale~\cite{luminosity-paper}. 
The integrated luminosity of the sample used in this analysis is determined to be 
$\mathcal{L}_{\rm int}= 34.7 \pm 1.2\invpb$, with an uncertainty
dominated by the knowledge of the beam currents. 

Only \X candidates for which the $J/\psi$ triggered the event are considered, keeping 70\% of the raw signal yield used for the mass measurement. 
In addition, the candidates are required to lie
inside the fiducial region for the measurement,
\begin{equation}
2.5 < y < 4.5 ~~~ \mbox{and} ~~~ 5 < p_{\rm T} < 20\gevc \,,
\label{eq:fiducial}
\end{equation}
where $y$ and $p_{\rm T}$ 
are the rapidity and transverse momentum of the \X. 
This region provides a good balance between a high efficiency (92\% of the triggered events)
and a low systematic uncertainty on the acceptance correction. 

The corrected yield $N^{\rm corr}_{\X} = 9140 \pm 2224$ is obtained from a mass fit
in the narrow region $3820-3950\mevcc$, with a linear background model and the same \X signal model as used
previously but with the mass and resolution fixed to the central
values presented in Sect.~\ref{sec:mass}. In this fit, each 
candidate is given a weight equal to the reciprocal of the 
total signal efficiency estimated from simulation for the
$y$ and $p_{\rm T}$ of that candidate. A second method based on the 
sWeight~\cite{2005NIMPA.555..356P} technique was found to give
consistent results. 
The average total signal efficiency in the fiducial region of Eq.~\ref{eq:fiducial}
is estimated to be $N_{\X}/N^{\rm corr}_{\X} = 4.2\%$, where $N_{\X}$ is the observed signal 
yield obtained from a mass fit without weighting the events. This low value of the efficiency is driven by the geometrical acceptance and the requirement on the $p_{\rm T}$ of the $J/\psi$ meson.

\begin{table}[t]
\caption{\small Relative systematic uncertainties on the \X production
cross-section measurement. The total uncertainty is the quadratic
sum of the individual contributions.}
\begin{center}
~ \\[-1ex]
\begin{tabular}{l|c} 
Source of uncertainty & $\Delta\sigma/\sigma$ [\%] \\ \hline 
\X polarization &   $2.1$ \\
\X decay model & $1.0$ \\
\X decay width & $5.0$ \\
Mass resolution & $2.5$ \\
Background model & $6.4$ \\
Tracking efficiency & 7.4 \\
Track $\chi^2$ cut & 2.0 \\
Vertex $\chi^2$ cut & 3.0 \\
Muon trigger efficiency & 2.9 \\
Hit-multiplicity cuts & 3.0  \\ 
Muon identification & 1.1 \\
Pion identification & 4.9 \\
Integrated luminosity & 3.5 \\
$J/\psi\to\mu^+\mu^-$ branching fraction & 1.0  \\ 
\hline
Total &  $14.2$ \\
\end{tabular}
\end{center}
\label{tab:syst}
\end{table}

The quantity $\xi$ of Eq.~\ref{eq:sigmaBR} is the product of three factors. The first two, 
$1.024 \pm 0.011$~\cite{LHCb_Jpsi_production}
and $0.869 \pm 0.043$, 
account for differences between the data and simulation for
the efficiency of
the muon and pion identifications, respectively.
The third factor, $0.92 \pm 0.03$, corresponds to the efficiency
of the hit-multiplicity cuts applied in the trigger, 
which is not accounted for in the simulation.
It is obtained from a fit of the distribution of the number of hits in the VELO.

The relative systematic uncertainties assigned to the cross-section 
measurement are listed in Table~\ref{tab:syst}, and quadratically add up to 14.2\%. 
The cross-section measurement is performed under
the most favoured assumption for the quantum numbers of the \X particle, $J^{PC}= 1^{++}$~\cite{Brambilla:2010cs},
which is used for the generation of Monte Carlo events. No systematic uncertainty is assigned to cover other cases. 
Besides the uncertainties already mentioned on $\mathcal{B}(J/\psi\to\mu^+\mu^-)$,
$\mathcal{L}_{\rm int}$ and $\xi$, the following sources of
systematics on $N^{\rm corr}_{\X}$ are considered. The dominant uncertainty is due to differences in the
efficiency of track reconstruction between the data and
simulation. This is estimated to be $7.4\%$ using
a data driven tag and probe approach based on 
$J/\psi \rightarrow \mu^+ \mu^-$ candidates. An additional uncertainty of 0.5\% per
track is assigned  to cover differences in the efficiency of the track
$\chi^2/{\rm ndf}$ cut between data and simulation.  
Similarly, a 3\% uncertainty is assigned due to the effect of the vertex $\chi^2$ cuts.

Other important sources of uncertainty are due to the modelling
of the signal and background mass distributions.
Repeating the mass fit with the \X decay width fixed to 
$0.7~\mevcc$ instead of zero results in a 5\% change of the 
signal yield.
Similarly, the uncertainties due to the \X mass resolution are estimated 
by repeating the mass fit with different fixed mass resolutions:
first changing it by the statistical uncertainty reported in Table~\ref{tab:fitResults},
and then changing it by the systematic uncertainty resulting from the knowledge of the resolution ratio 
$\sigma_{\X}/\sigma_{\psi(2S)}$, as described in Sect.~\ref{sec:mass}.
The combined effect on the $X(3872)$ signal yield corresponds to a 2.5\% systematic uncertainty. 

Using an exponential rather than linear function to describe the background
leads to a change of 6.4\% in signal yield, which 
is taken as an additional systematic uncertainty.

The unknown \X polarization affects the total efficiency,
mainly through the $J/\psi$ reconstruction efficiency. 
The dipion system is less affected, in particular the
efficiency is found to be constant as a function of the dipion mass. 
The simulation efficiency, determined assuming no $J/\psi$ polarization, 
is recomputed in two extreme schemes for the $J/\psi$ polarization
(fully transverse and fully longitudinal)~\cite{LHCb_Jpsi_production}
and the maximum change of 2.1\% is taken as systematic uncertainty. 
The efficiency of the $Q$ cut depends on the \X decay model. The
dipion mass spectrum obtained in this analysis does not have enough accuracy to
discriminate between reasonable models. Comparing the results obtained
with the $\X \to J/\psi \rho$ decay models used by CDF~\cite{Abulencia:2005zc}  
and by Belle~\cite{Choi:2011fc}, we evaluated a 1\% systematic uncertainty on the $Q$-cut efficiency. 

Finally, differences in the trigger efficiency
between data and simulation are studied using events
triggered independently of the $J/\psi$ candidate. Based on these
studies an uncertainty of 2.9\% is assigned.

\section{Results and conclusion}
\label{summary}
With an integrated luminosity of 34.7\invpb 
collected by the LHCb experiment, 
the production of the \X particle
is observed in $pp$ collisions at $\sqrt{s}= 7\tev$.
The product of the production cross-section and the branching ratio into $J/\psi\pi^+\pi^-$ is
\begin{equation}
\sigma(pp\to \X+{\rm anything}) \, \mathcal{B}(\X \to J/\psi\pi^+\pi^-) = 
\theresult \pm \thestat \,{\rm (stat)} \pm \thesyst \,{\rm (syst)}\nb \,, \nonumber
\end{equation} 
for \X mesons produced (either promptly or from the decay of other particles) 
with a rapidity between 2.5 and 4.5 and a transverse momentum between
5 and 20\gevc.

Predictions for the $\X\to J/\psi \pi^+\pi^-$
production at the LHC are available from a
non-relativistic QCD model which assumes that the cross-section is
dominated by the production of
charm quark pairs with negligible relative momentum~\cite{Artoisenet:2009wk}. 
The calculations are normalized using extrapolations from measurements performed at the Tevatron. When restricted to the kinematic range of 
our measurement and summed over prompt production and production from 
$b$-hadron decays, the results of Ref.~\cite{Artoisenet:2009wk} yield
$13.0 \pm 2.7\nb$, where the quoted uncertainty originates from the 
experimental inputs used in the calculation.
This prediction exceeds our measurement by $2.4\sigma$. 

After calibration using $J/\psi\to\mu^+ \mu^-$ decays, the masses
of both the \X and $\psi(2S)$ mesons, 
reconstructed in the same $ J/\psi \pi^+ \pi^-$ final state, 
are measured to be
\begin{eqnarray*}
m_{\X} &=& 3871.95 \pm 0.48 \,({\rm stat}) \pm 0.12 \,({\rm syst})\mevcc \,, \\
m_{\psi(2S)} &=& 3686.12\pm 0.06 \,({\rm stat}) \pm 0.10 \,({\rm syst})\mevcc \,,
\end{eqnarray*}
in agreement with the current world averages~\cite{Nakamura:2010zzi},
and with the recent 
\X mass measurement from Belle~\cite{Choi:2011fc}. The measurements of
the \X mass are consistent, within uncertainties, with the sum of the $D^0$ and $D^{*0}$ masses,
$3871.79 \pm 0.29~\mevcc$, computed 
from the results of the global PDG fit of the charm meson masses~\cite{Nakamura:2010zzi}.

\section*{Acknowledgements}

We thank P.\ Artoisenet and E.\ Braaten for useful discussions and for recomputing the numerical
prediction of Ref.~\cite{Artoisenet:2009wk} in the fiducial region of our measurement.
We express our gratitude to our colleagues in the CERN accelerator
departments for the excellent performance of the LHC. We thank the
technical and administrative staff at CERN and at the LHCb institutes,
and acknowledge support from the National Agencies: CAPES, CNPq,
FAPERJ and FINEP (Brazil); CERN; NSFC (China); CNRS/IN2P3 (France);
BMBF, DFG, HGF and MPG (Germany); SFI (Ireland); INFN (Italy); FOM and
NWO (The Netherlands); SCSR (Poland); ANCS (Romania); MinES of Russia and
Rosatom (Russia); MICINN, XuntaGal and GENCAT (Spain); SNSF and SER
(Switzerland); NAS Ukraine (Ukraine); STFC (United Kingdom); NSF
(USA). We also acknowledge the support received from the ERC under FP7
and the Region Auvergne.

\bibliographystyle{LHCb}
\bibliography{refs}

\end{document}